\documentclass[10pt,a4]{article}
\usepackage[a4paper,bottom=1in,top=1in,left=1in,right=1in]{geometry}
\usepackage{graphicx}
\usepackage{times} 
\usepackage{graphicx} 
\usepackage{dcolumn}
\usepackage{bm}
\usepackage{amssymb}
\usepackage{amsmath}
\usepackage{wasysym}
\usepackage{color}
\usepackage{hyperref}
\usepackage[normalem]{ulem}
\usepackage{float}
\usepackage{flushend}
\usepackage{multirow}
\usepackage{cancel}
\usepackage{cite}
\usepackage[dvipsnames]{xcolor}
\hypersetup{
colorlinks,
citecolor=blue,
filecolor=blue,
linkcolor=blue,
urlcolor=blue}

\title{Turbulence in a stably stratified fluid: Onset of global anisotropy as a function of the Richardson number}

\author{Jayanta~K.~Bhattacharjee$^{1,}$\thanks{Email: jayanta.bhattacharjee@gmail.com}, Abhishek Kumar$^{2,}$\thanks{Email: abhishek.kir@gmail.com}, and Mahendra K. Verma$^{3,}$\thanks{Email: mkv@iitk.ac.in}\\
$^1$Department of Theoretical Physics,\\
Indian Association for the Cultivation of Science, \\
Kolkata 700 032, India\\
$^2$ Centre for Fluid and Complex Systems,\\
Coventry University, Coventry CV1 5FB, UK\\
$^3$Department of Physics,\\
Indian Institute of Technology, Kanpur 208016, India}

\date{}   

\begin{document}
\maketitle
\abstract{It is necessary to introduce an external forcing to induce turbulence in a stably stratified fluid. The Heisenberg eddy viscosity technique should in this case suffice to calculate a space-time averaged quantity like the global anisotropy parameter as a function of the Richardson number. We find analytically that the anisotropy  increases linearly with the Richardson number, with a small quadratic correction.  A numerical simulation of the complete equations shows the linear behaviour.}

\section{Introduction}
\label{sec:intro}
For the study of homogeneous isotropic turbulence, the velocity structure factors play a very important role since they probe the probability distribution function of the turbulent velocity field. While experiments and numerical simulations can study the probability distribution directly, it is far more difficult to calculate it.  Hence analytic calculations generally deal with the structure factors, and among them,  the two-point correlation function is of particular importance since it leads to the energy spectrum. In the case of turbulence in a stratified fluid, besides the structure factor, there is the anisotropy which is an essential  characteristic. In this article, we consider a stably stratified fluid and the global anisotropy parameter which is a ratio of the kinetic energy in the plane perpendicular to the stratification direction to the kinetic energy in the vertical direction. We will restrict ourselves to the case of weak turbulence where the Fourier space structure factors are still determined by the magnitude of the wavenumber $k$, where $k^2 = k^2_{\perp} + k^2_{\parallel}$. The subscripts $\perp$ and $\parallel$ refers to the horizontal plane and the vertical direction respectively.  It is our contention that for the stably stratified fluid where an external forcing is necessary to induce turbulence, the equivalent linearization will be useful to calculate a global quantity. We will consequently use this technique to calculate the anisotropy in perturbation theory and compare it with  a numerical simulation.

One of the cornerstones of the theory of homogeneous isotropic turbulence is Kolmogorov's $-5/3$ law~\cite{Kolmogorov:DANS1941Structure}. If the total kinetic energy per unit mass is $E$, then its distribution $E(k)$ over the different wavenumbers ($k$ is the inverse of the length scale $l$) is defined by
\begin{equation}
E = \int E(k)dk.
\end{equation}
In the  inertial range, i.e., a range of wavenumbers much smaller than the viscous dissipation wavenumber and at the same time much larger than the wavenumber corresponding to the macroscopic energy input scale, Kolmogorov motivated a dimensional analysis to write
\begin{equation}
E(k) = C \epsilon^{2/3} k^{-5/3}, \label{eq:Ek_Kolmo}
\end{equation}
where $\epsilon$ is the rate at which energy is dissipated by viscosity, and $C$ is the Kolmogorov's constant. In the stationary state, which is the case being considered, $\epsilon$ equals the rate at which the energy is injected into the system. Behind Kolmogorov's analysis lies the idea of a constant energy cascade from large length scales (small wavenumbers) to short length scales (large wavenumbers) brought about by the nonlinear terms in the Navier--Stokes equation. The rate of the cascade is the same as $\epsilon$ that has been introduced earlier.

Turbulence in a stratified fluid offers additional possibilities. The isotropy is now broken by a density gradient in a particular direction (we will take this as the $z$-direction which is also the direction in which gravity acts). For our analysis, we consider a thermal gradient produced by a constant temperature difference maintained across two parallel plates separated by a distance $d$ in the $z$-direction. The density gradient produced creates buoyancy forces which play a strong role. The problem now has two coupled fields---the velocity field ${\bf v}({\bf r},t)$, and the temperature (scalar) field $T({\bf r},t)$. It was argued by Bolgiano~\cite{Bolgiano:JGR1959} and independently by Obukhov~\cite{Obukhov:DANS1959} that under these circumstance, there would be two fluxes---one for the usual kinetic energy, and the other for the temperature fluctuation $\delta T({\bf r},t)$ which is the deviation of temperature $T({\bf r},t)$ from the steady state profile of $T_s (z) = T_1 + \Delta T z/d $. Here $T_1$ is the temperature of the lower plate $(z=0)$, $T_2$ the temperature of the upper plate $(z=d)$, and $\Delta T = T_2 - T_1 > 0$. The plates are perfect conductors of heat. Analogous to the  kinetic energy $E$, we define { a thermal energy} as  
\begin{equation}
G = \frac{1}{2V} \int  (\delta T)^2 d{\bf r},
\label{eq:G}
\end{equation}
where $V$ is the volume of the fluid. It is a conserved quantity in the absence of external forcing, buoyancy, and thermal diffusion that dissipates the fluctuations at short length scales. It was suggested by Bolgiano~\cite{Bolgiano:JGR1959}  and Obukhov~\cite{Obukhov:DANS1959} that there would be a cascade of $G$ in the event of a buoyancy-driven turbulence, and the rate of dissipation of $G$ in the steady state would equal the rate of input at large scales with the constant being the {thermal} flux $\chi = dG/dt$. Dimensional considerations similar to that of Kolmogorov yields~\cite{Bolgiano:JGR1959,Obukhov:DANS1959}
\begin{equation}
E(k) = \tilde C \chi^{2/5} (\alpha g)^{4/5} k^{-11/5}, \label{eq:Ek_Bol}
\end{equation}
where $\tilde C$ is a number of $\mathcal{O}(1)$, $\alpha$ is the expansion coefficient and $g$ the acceleration due to
gravity. This is called the Bolgiano--Obukhov scaling.

Despite several numerical and experimental efforts~\cite{Kimura:JFM1996,WAITE:JFM19992,Lindborg:JFM2006,Brethouwer:JFM2007,Vallgren:PRL2011,Bartello:JFM2013,Kimura:JFM2012,Zhang:PRL2005,Seychelles:PRL2008} over a decade, a clear observation of the exponent $11/5$ was not seen until the numerical work of Kumar {\em et. al.}~\cite{Kumar:PRE2014}. They showed a clear $11/5$ spectrum,  a corresponding flat {thermal} flux, and a $k$-dependent kinetic energy flux when the fluid is heated from above, i.e., for a stably stratified fluid. Note however that Kumar {\em et al.}~\cite{Kumar:PRE2014} observed $k^{-11/5}$ kinetic-energy spectrum for the nearly isotropic regime of stably-stratified turbulence.  The behaviour is quite different for quasi two-dimensional stably-stratified turbulence~~\cite{Kimura:JFM1996,Lindborg:JFM2006,Brethouwer:JFM2007,Vallgren:PRL2011,Bartello:JFM2013,Kimura:JFM2012,Zhang:PRL2005,Seychelles:PRL2008,deBruynKops:JFM2015}.   
The validity of Eq.~(\ref{eq:Ek_Bol}) clearly requires the dominance of the thermal flux and this is possible for a large Richardson number $\mathrm{Ri}$, which is defined as~\cite{Davidson:book:TurbulenceRotating,Kumar:PRE2014,Bhattacharjee:PLA2015}  
\begin{equation}
\mathrm{Ri} = \frac{\alpha g | \Delta T| d}{v^{2}_s},
\end{equation}
where $v_s$ is the rms of the velocity. The numerator is a measure of the square of the velocity of free fall under gravity. If the inherent fluctuations (in the absence of buoyancy, or for weak buoyancy) dominate, then the Richardson number will be very small, and we expect the Kolmogorov spectrum.  Our interest in this work is to look at small values of $\mathrm{Ri}$ and see how the anisotropy sets in as the temperature gradient is increased.  

The anisotropy should be apparent from the study of the spectrum. However, the spectrum is a local quantity and difficult to measure. A global quantity could be an alternative measure for studying the crossover between the Kolmogorov and Bolgiano-Obukhov regimes. The global anisotropy
\begin{equation}
A =  \frac{\int \bar{v}^2_\perp d {\bf r}}{2 \int \bar{v}^2_\parallel d {\bf r}} =  \frac{\int (\bar{v}^2_1 + \bar{v}^2_2) d {\bf r}}{2 \int \bar{v}^2_3 d {\bf r}}
\label{eq:aniso_def}
\end{equation}
could be such a parameter. Here $v_{\parallel}$ is the velocity component in the $z$-direction, i.e., parallel to the gradient, while $v_{1,2}$ are the velocity components in the $x$-$y$ plane and $\bar{v}^2_\perp = \bar{v}^2_1 + \bar{v}^2_2$ . The overbars indicate the time-averaged value. 

Quantification of anisotropy in the stably stratified flow has been remaining a significant topic of interest. Kaneda and Yoshida~\cite{Kaneda:NJP2004}  expressed the velocity correlation function using tensors and quantified the small-scale anisotropy in stably stratified turbulence. Later, Ishida and Kaneda~\cite{Ishida:PF2007} applied the same for the quasi-static magnetohydrodynamic turbulence~\cite{Verma:ROPP2017}. Rorai {\em et al.}~\cite{Rorai:PRE2015} computed the axisymmetric kinetic energy spectrum to quantify anisotropy in the wavenumber space for the stably stratified turbulence.

The outline of the paper is as follows. In Sec.~\ref{sec:theory}, we calculate the global anisotropy ratio, $A$, as a function of the Richardson number using an {\em equivalent linearization technique} based on the Heisenberg picture of eddy viscosity. In Sec.~\ref{sec:numerical}, we perform numerical simulations of stably stratified turbulence and test  our expression for the global anisotropy obtained in Sec.~\ref{sec:theory}.  Finally, we conclude in Sec.~\ref{sec:conc}.

\section{Theory}
\label{sec:theory}
The governing equations are the incompressible Navier--Stokes equation and heat conduction equation with random forces. We  consider a stable stratification, i.e., the temperature at the top plate is larger than that of the bottom plate, $T_2 > T_1$ with $\Delta T$ positive. Since this system is stable, we need external forcing to maintain a steady-state turbulence. We will use a Boussinesq approximation~\cite{Chandrasekhar:book:Instability} where the density fluctuation appears only in the buoyancy, and  will treat the fluid as incompressible. Consequently, the Navier--Stokes equation reads (as used in fluctuating hydrodynamics)
\begin{equation}
\frac{\partial {\bf v}}{\partial t} +  (\mathbf{v} \cdot \nabla) {\bf v}  = -  \frac{\nabla \delta p}{\rho_0} +\nu \nabla^2{\bf v}+  \alpha g \delta T  \delta_{i3} + {\bf f}^\prime, \label{eq:u}
\end{equation}
\begin{equation}
\nabla \cdot {\bf v}  = 0,
\end{equation}
with $\nabla \cdot {\bf f}^\prime = 0$. In Eq.~(\ref{eq:u}), $\delta p$ is the pressure fluctuation about the steady conduction state, $\nu$ is the kinematic viscosity, $\rho_0$ is the background density, and ${\bf f}^\prime$ the fluctuating force. The heat conduction equation reads,
\begin{equation}
\frac{\partial}{\partial t} \delta T + (\mathbf{v} \cdot \nabla) \delta T = \lambda \nabla^2\delta T - \frac{\Delta T}{d}v_3  + h^\prime, \label{eq:T}
\end{equation}
where $\delta T$ is the temperature fluctuation form the conduction state, $\lambda$ is the thermal diffusion coefficient, and $h^\prime$ is the fluctuating force. Usually one does not consider a fluctuating source term in Eq.~(\ref{eq:T}). For generality, however, we include a random source term $h^\prime$ in Eq.~(\ref{eq:T}) as well.

We now carry out the  rescaling of Eqs.~(\ref{eq:u})-(\ref{eq:T}), using $d$ as the length scale, the rms value of the fluctuating turbulent velocity field $v_s$ as the velocity scale, $d/v_s$ as the time scale, $\Delta T$ as the temperature scale,  which leads to:
\begin{itemize}
\item  ${\bf v}={\bf u} v_s$;
\item $\delta T = \theta \Delta T$;
\item $t = \tau d/v_s$;
\item ${\bf f}^\prime={\bf f} v_s^2/d$; ${h}^\prime={h} v_s^2/d$.
\end{itemize}
This unconventional rescaling will make the crossover with changing Ri apparent. The pressure fluctuations and density are then appropriately scaled to make the resulting system appear as follows: 
\begin{eqnarray}
\frac{\partial \mathbf{u}}{\partial \tau} + (\mathbf{u} \cdot \nabla) \mathbf{u} & = & - \frac{\nabla \delta p}{\rho_0} + \mathrm{Ri}\theta \hat{\mathbf{z}} + \frac{\nu}{v_s d} \nabla^2 \mathbf{u} + {\bf f} \label{eq:u_non}, \\
\nabla \cdot {\bf u} & = &0 \label{eq:inc_non},\\
\frac{\partial \theta}{\partial \tau} + (\mathbf{u} \cdot \nabla) \theta & = & \frac{\lambda}{v_s d} \nabla^2 \theta - u_3 + h. \label{eq:theta_non}
\end{eqnarray}
The recast has made the crossover apparent. For ${\mathrm{Ri} \rightarrow 0} $, the velocity dynamics is independent of the temperature fluctuations, and the temperature slaves to the velocity field. This is the Kolmogorov limit. The thermal fluctuations begin to affect the system as $\mathrm{Ri}$ increases, and the crossover takes place. The fluctuating {force ${\bf f}$ and current $h$} are Gaussian random fields and are specified by the correlations
\begin{eqnarray}
\langle f_i (\mathbf{k},\tau) f_j (\mathbf{k^\prime},\tau^\prime) \rangle  =  P_{ij}(\mathbf{k}) \delta (\mathbf{k} + \mathbf{k^\prime}) \delta (\tau-\tau^\prime) F(k), \label{eq:ff}\\
\langle h (\mathbf{k},\tau) h (\mathbf{k^\prime},\tau^\prime) \rangle  =  \delta (\mathbf{k} + \mathbf{k^\prime}) \delta (\tau-\tau^\prime) H(k), 
\end{eqnarray}
where $P_{ij}(k)$ is the projection operator, ${\bf k}$ is the wave-vector, and $F(k)$ and $H(k)$ are functions which need to be compatible with the scale invariance of Eqs.~(\ref{eq:u_non})-(\ref{eq:theta_non}). Scale invariant forces $F(k)$ and $H(k)$ to be proportional to $k^{-D}$, where $D$ is the dimensionality of space. It is easily seen from Eqs.~(\ref{eq:u_non})-(\ref{eq:theta_non}) that energy and entropy conservations have the form
\begin{eqnarray}
 \frac{\partial}{\partial \tau} \int  \frac{u^2}{2} d{\bf r}   =  \int  \left( \mathrm{Ri}\theta u_3 + f_i u_i + \nu^\prime (\partial_i u_i)^2 \right) d{\bf r},\\
\frac{\partial}{\partial \tau} \int  \frac{\theta^2}{2} d{\bf r}  =  \int  \left( -\theta u_3 + \theta h + \lambda^\prime (\partial_i \theta)^2 \right) d{\bf r},
\end{eqnarray}
which leads to
\begin{eqnarray}
\frac{\partial}{\partial \tau} \int  \left[ \frac{u^2}{2}  +   \frac{\mathrm{Ri} \theta^2}{2} \right] d{\bf r}   = \int  \left( \nu^\prime(\partial_i u_i)^2 + \lambda \mathrm{Ri}(\partial_i \theta)^2 + f_i u_i + \mathrm{Ri}h \theta \right) d{\bf r}
\end{eqnarray}
showing clearly that the conserved quantity in the ``no-forcing", ``no-dissipation" limit is $(u^2+\mathrm{Ri}\theta^2)/2$, consistent with the crossover picture.

We now turn to the strategy for calculating the anisotropy factor $A$. We adopt a familiar technique of
nonlinear dynamics, {\em equivalent linearization}. This is precisely what is meant by Heisenberg's eddy viscosity~\cite{Heisenberg:PRA1948,Chandrasekhar:PRSA1949}. It was argued by Heisenberg~\cite{Heisenberg:PRA1948} and later amplified by Chandrasekhar~\cite{Chandrasekhar:PRSA1949} that the effect of the nonlinear term in the inertial range is to transfer energy from small-$k$ to large-$k$ exactly as molecular viscosity would do. However, this effective viscosity is scale dependent and proportional to $l^{4/3}$ and dominates the contribution from molecular viscosity. Consequently, we replace the nonlinear term by an effective viscous term $\nu_\mathrm{eff}(k)k^2$ and drop the molecular viscosity contribution {(the technical details are in \ref{sec:appA})}. Since $\nu_\mathrm{eff}(k) \propto k^{-4/3}$, the relaxation rate is  $\nu_\mathrm{eff}(k)k^2 = \Gamma_1 k^{2/3}$. Since the $\theta$-dynamics slaves to the velocity field, there is identical scaling for the thermal diffusivity which makes the associated relaxation rate $\lambda_\mathrm{eff}(k)k^2 = \Gamma_2 k^{2/3}$ . However, before implementing the procedure, we need to eliminate $\delta p$ from Eq.~(\ref{eq:u_non}). This is done by taking the divergence of Eq.~(\ref{eq:u_non}) to obtain formally
\begin{equation}
-\nabla^2 \left( \frac{\delta p}{\rho_0}\right) = \nabla \cdot \left[ (\mathbf{u} \cdot \nabla) \mathbf{u} \right] - \mathrm{Ri} \frac{\partial \theta}{\partial z}. \label{eq:lab_pressure}
\end{equation}
Using Eq.~(\ref{eq:lab_pressure}) in  Eqs.~(\ref{eq:u_non})-(\ref{eq:theta_non}) along with the above equivalent relaxations, we have in wave-vector space
\begin{eqnarray}
\frac{\partial u_\alpha ({\bf k})}{\partial \tau} & = & - \Gamma_1 k^{s} {u}_\alpha ({\bf k}) \nonumber \\
& + & \mathrm{Ri} \left( \delta_{\alpha 3} - \frac{k_\alpha k_3}{k^2} \right) \theta({\bf k})  + f_\alpha({\bf k}) , \label{eq:u_alpha_k}\\
\frac{\partial \theta ({\bf k})}{\partial \tau} & = & -\Gamma_2 {k}^{s} \theta ({\bf k}) - {u}_3 ({\bf k}) + h ({\bf k}). \label{eq:theta_k}
\end{eqnarray}
In the above $s$ is an exponent which will later be set equal to $2/3$ in $D=3$. The horizontal component of $\bf k$ form a continuum, while the vertical component is discrete (due to boundary conditions at $z=0$ and $z=d$) and the allowed values are $n\pi$, where $n$ is a positive integer. {In the above calculation, we have assumed in Eqs.(\ref{eq:u_alpha_k}) and (\ref{eq:theta_k}) that the effective relaxation rate $\Gamma_{1,2}k^s$ depend only on the magnitude of the wavevector $k=\sqrt{k_1^2+k_2^2+n^2\pi^2/d^2}$. This is allowed for small Richardsons number where the flow is nearly isotropic. Here we study the onset of anisotropy.  Note that the flow becomes strongly anisotropic only at large Richardson number~\cite{PRAUD:JFM2005}.}

Working in frequency space {with the Fourier transform defined as 
\begin{equation}
\phi(t)=\int \frac{d\omega}{2\pi}e^{-i\omega t}\phi(\omega)
\end{equation}
 for any function $\phi(t)$}, we find [where $\xi =  \cos^{-1}({k_3/{k}})$] 
\begin{eqnarray}
u_3 & = & \frac{f_3 \left( -i \omega + \Gamma_2 k^{s}  \right) + h\mathrm{Ri} \sin^2 \xi}{\Delta},\\
u_\alpha & = & \frac{f_\alpha}{-i \omega + k^{s}} + \frac{f_3 \mathrm{Ri} k_\alpha k_3}{k^2 \left( -i \omega + \Gamma_1 k^{s} \right)\Delta} \nonumber \\
& - & \frac{\mathrm{Ri} k_\alpha k_3 h}{k^2 \Delta};\, \alpha = 1\, {\rm or}\,2,
\end{eqnarray}
with
\begin{eqnarray}
\Delta & = & \left( -i \omega + \Gamma_1 k^{s} \right) \left( -i \omega + \Gamma_2 k^{s} \right) + \mathrm{Ri} \sin^2 \xi.
\end{eqnarray}
{It should be noted that in the absence of the {external current} `$h$', the vertical velocity goes to zero as $\mathrm{Ri} \rightarrow \infty$. This means that the flow will become two-dimensional at large Richardson number.} The time averaged values are obtained as $\int |u_3|^2 d \omega/2 \pi$ and $ \int \left( |u_1|^2 + |u_2|^2 \right) d \omega/2\pi $.  After performing relevant contour integrals, we obtain
\begin{equation}
\bar{u_3}^2  =  \frac{1}{2 \pi} \int |u_3|^2 d\omega \nonumber
\end{equation}
\begin{equation}
=\frac{\left( \Gamma_2 \left( \Gamma_1 + \Gamma_2 \right) k^{2{s}} + \mathrm{Ri} \sin^2 \xi \right) \langle f_3f_3 \rangle + 2\mathrm{Ri}^2 \sin^4 \xi \langle hh \rangle}{2\left( \Gamma_1 + \Gamma_2 \right)k^{s} \left( \Gamma_1 \Gamma_2 k^{2{s}} + \mathrm{Ri} \sin^2 \xi \right)} \label{eq:u_3_bar}.
\end{equation}
Similarly we have
\begin{eqnarray}
\bar{u_1}^2 + \bar{u_2}^2 & = & \frac{1}{2\pi} \int \left( |u_1|^2 + |u_2|^2 \right) d\omega  \nonumber \\ 
& = & \frac{\langle f_1 f_1 \rangle + \langle f_2 f_2 \rangle}{2 \Gamma_1 k^{s}} \nonumber \\ 
& + & \frac{\mathrm{Ri}^2 \cos^2 \xi \sin^2 \xi \left( 2 \Gamma_1 + \Gamma_2 \right) k^{s} \langle f_3 f_3 \rangle}{2 \Gamma_1 \left( \Gamma_1 + \Gamma_2 \right) k^{2{s}} \left( \mathrm{Ri} \sin^2 \xi + \Gamma_1 \Gamma_2 k^{2{s}} \right) \left( 2 \left( \Gamma_1 + \Gamma_2 \right) \Gamma_1 k^{2{s}} + \mathrm{Ri} \sin^2 \xi \right)} \nonumber \\ 
& + & \frac{\mathrm{Ri}^2 \cos^2 \xi \sin^2 \xi \langle hh \rangle}{\left( \Gamma_1 + \Gamma_2 \right) k^{s} \left( \Gamma_1 \Gamma_2 k^{2{s}} + \mathrm{Ri} \sin^2 \xi \right)}. \label{eq:u_12_bar}
\end{eqnarray}
Since Eq.~(\ref{eq:ff}) implies 
\begin{equation}
\langle f_1 f_1 \rangle = \langle f_2 f_2 \rangle = \langle f_3 f_3 \rangle = \langle f f \rangle,
\end{equation}
we see from Eqs.~[(\ref{eq:u_3_bar}),~(\ref{eq:u_12_bar})], that $A=1$ if $\mathrm{Ri} = 0$.

Having written out the general structure of the time averaged kinetic energies in the vertical and horizontal directions, we will drop the forcing in the temperature dynamics to compare with the numerical work where a forcing was used only in the velocity dynamics. To calculate the anisotropy factor explicitly, we need to do an integration over the wave-vectors in the horizontal direction and a sum over the discrete wavenumbers in the vertical direction.

The first correction to the isotropic limit (i.e., $\mathrm{Ri}=0$) is linear in $\mathrm{Ri}$ :
\begin{equation}
2 \bar{u_3}^2  =  \frac{\langle f f \rangle}{\Gamma_1 k^{s}} \left[ 1 - \mathrm{Ri} \frac{\left( {k_1}^2 + {k_2}^2 \right)}{k^{2+2{s}} \Gamma_1 \left( \Gamma_1 + \Gamma_2 \right)} + \mathcal{O}(\mathrm{Ri}^2) \right]
\end{equation}
\begin{equation}
\bar{u_1}^2 + \bar{u_2}^2  =  \frac{\langle f f \rangle}{\Gamma_1 k^{s}} \left[ 1 + \mathcal{O}(\mathrm{Ri}^2) \right].
\end{equation}
Here we  employ $\sin^2 \xi = (k_1^2 + k_2^2)/k^2$. The anisotropy factor  works out following an integration over the wavenumbers (with $\langle ff \rangle = F_0 /k^D$, where $F_0$ is constant). In  \ref{sec:appB}, we have calculated 
\begin{equation}
A = 1 + 0.067X + 0.016 X^2,
\label{eq:A_3D}
\end{equation}
where 
\begin{equation}
X=\frac{\mathrm{Ri}}{\Gamma_1 \Gamma_2 \pi^{4/3}}.
\end{equation}

\section{Numerical Simulation}
\label{sec:numerical}

To verify our approximate analytic results,  we performed  numerical simulations of stably stratified turbulence in both square and cubical geometries. We solve Eqs.~(\ref{eq:u})-(\ref{eq:T}) using a pseudo-spectral code Tarang~\cite{Chatterjee:JPDC17} with $h^\prime=0$. We employ periodic boundary conditions on all sides of the box for both the velocity field {\bf v} and the thermal fluctuations $\delta T$. We employ a fourth-order Runge--Kutta method for time stepping, the Courant--Friedrichs--Lewy condition to determine the time step $\Delta t$, and $2/3$ rule for dealiasing. For the two-dimensional $(D=2)$  simulation, the dimension of the square box is $(2\pi)^2$, while for the three-dimensional $(D=3)$ simulation, the size of the cubical box is $(2\pi)^3$.

{ For the numerical simulation, we turn off the fluctuating force $h^\prime$, and in order to obtain a steady stably-stratified turbulent flow, we apply a random force $\bf f(\bf k)$ in the wavenumber band $2 \leq k \leq 4$ to the velocity field using the following scheme:
\[ 
{{\bf f}}({\bf k}) = \Lambda_1
\left( \begin{array}{c}
\sin \phi \\
 - \cos \phi \\
0
\end{array} \right) + \Lambda_2
\left( \begin{array}{c}
\cos \vartheta \cos \phi \\
 \cos \vartheta \sin \phi \\
- \sin \vartheta
\end{array} \right),
\] 
where $\vartheta$ and $\phi$ are usual polar and azimuthal angles, respectively. $\Lambda_1$ and $\Lambda_2$ are the product of random phase and the forcing amplitude $\mathcal{A}$ given by following formula:
\begin{eqnarray}
\Lambda_1 & = & \mathcal{A} \exp(i \Phi_1) \cos \Phi_3, \\
\Lambda_2 & = & \mathcal{A} \exp(i \Phi_2) \sin \Phi_3,
\end{eqnarray}
and 
\begin{equation}
\mathcal{A} = \sqrt{\frac{2\varepsilon}{n_f\Delta t}}.
\end{equation}
Here $\Phi_j$'s ($j=1,2,3$) follow uniform distribution in $[0, 2\pi]$ with zero mean, $\varepsilon$ is the constant energy supply rate to the system, and $n_f$ is the total number of modes inside the forcing wavenumber band. }

We nondimensionalize Eqs.~(\ref{eq:u})-(\ref{eq:T}) such that the Rayleigh number $\mathrm{Ra}$, the Prandtl number $\mathrm{Pr}$, and the  total energy supply rate $\varepsilon$ of the external forcing are the control parameters. The Rayleigh number $\mathrm{Ra}$ is the ratio of the buoyancy and the viscous force,  defined as 
\begin{equation}
\mathrm{Ra} = \frac{\alpha g \Delta T d^3}{\nu \lambda},
\end{equation}
and the Prandtl number $\mathrm{Pr} = \nu/ \lambda$. Note that the Richardson number $\mathrm{Ri}$, the Froude number $\mathrm{Fr} \approx 1/\sqrt{\mathrm{Ri}}$~\cite{Verma:book}, and the Reynolds number $\mathrm{Re}$  are the response parameters. {We fix $\mathrm{Pr}=1$, and vary $\varepsilon$ and $\mathrm{Ra}$ to obtain adequate Richardson number. Note that, the Rayleigh number controls the buoyancy force, while $\varepsilon$ regulates the external forcing.}   For further details of the numerical simulation, we refer Kumar {\em et al.}~\cite{Kumar:PRE2014,Kumar:JoT2016} and {Verma} {\em et al.}~\cite{verma:NJP2017}.  We list all our parameters for the $D=3$ and $D=2$ simulations in Table~\ref{table:simulation_3D} and Table~\ref{table:simulation_2D} respectively.

\setlength{\tabcolsep}{5pt}
\begin{table}[htbp]
\begin{center}
\caption{Parameters of our direct numerical simulations (DNS) for three-dimensional stably stratified turbulence: Richardson number $\mathrm{Ri}$; grid resolution; Rayleigh number $\mathrm{Ra}$; energy supply rate $\varepsilon$; anisotropy ratio $A=\langle u_{\perp}^2 \rangle/(2\langle u_{\parallel}^2 \rangle) = \langle u_1^2 +u_2^2\rangle/(2 \langle u_3^2\rangle) $; Reynolds number $\mathrm{Re}$; Froude number $\mathrm{Fr}$; the kinetic energy dissipation rate $\epsilon$, the potential energy dissipation rate $\epsilon_{\chi}$; and $k_{\rm max}\eta$, where $k_{\rm max}$ is the maximum wavenumber and $\eta$ is the Kolmogorov length. For all our runs the Prandtl number $\mathrm{Pr}=1$.}
\vspace{5 mm}
\begin{tabular}{c  c c  c c c c c c c}
\hline \hline 
$\mathrm{Ri}$  &${\rm Grid}$ & $\mathrm{Ra}$   & $\varepsilon$  &$A$ &$\mathrm{Re}$ & $\mathrm{Fr}$& $\epsilon$ & $\epsilon_{\chi}$ & $k_{\rm max}\eta$ \\[1 mm]
\hline
$2.7$  & $512^3$ & $10^6$ & $0.1$ & $1.3$ & $1.4 \times 10^3$ & $0.6$ & $0.02$ & $0.02$ & $2.0$\\
$0.5$  & $512^3$ &  $10^5$ & $5$ & $1.2$ & $467$ & $1.4$ & $0.47$ & $60.7$ & $4.2$\\
$0.01$  & $1024^3$ &  $5 \times 10^3$ & $10^3$ & $1.0$ & $649$ & $10$ & $114$ & $150$ & $6.4$\\
\hline 
\end{tabular}
\label{table:simulation_3D}
\end{center}
\end{table}

\setlength{\tabcolsep}{4pt}
\begin{table}[htbp]
\begin{center}
\caption{Parameters of our direct numerical simulations (DNS) for two-dimensional stably stratified turbulence ($D=2$): Richardson number $\mathrm{Ri}$; grid resolution; Rayleigh number $\mathrm{Ra}$; energy supply rate $\varepsilon$; anisotropy ratio $A=\langle u_{\perp}^2 \rangle/\langle u_{\parallel}^2 \rangle  = \langle u_1^2 \rangle/ \langle u_3^2\rangle  $; Reynolds number $\mathrm{Re}$; Froude number $\mathrm{Fr}$; the kinetic energy dissipation rate $\epsilon$, the potential energy dissipation rate $\epsilon_{\chi}$; and $k_{\rm max}\eta$. For all our runs the Prandtl number $\mathrm{Pr}=1$.}
\vspace{5 mm}
\begin{tabular}{c  c c  c c c c c c c}
\hline \hline 
$\mathrm{Ri}$  &${\rm Grid}$ & $\mathrm{Ra}$   & $\varepsilon$  &$A$ &$\mathrm{Re}$ & $\mathrm{Fr}$& $\epsilon$ & $\epsilon_{\chi}$ & $k_{\rm max}\eta$ \\[1 mm]
\hline
$0.5$  & $2048^2$ & $10^8$ & $0.6$ & $1.1$ & $1.5 \times 10^4$ & $1.5$ & $1.3 \times 10^{-2}$ & $0.3$ & $3.0$\\
$0.8$  & $2048^2$ &  $10^8$ & $0.3$ & $1.4$ & $1.1 \times 10^4$ & $1.1$ & $9.2 \times 10^{-3}$ & $7.7 \times 10^{-2}$ & $3.3$\\
$1.9$  & $2048^2$ &  $10^8$ & $0.1$ & $1.6$ & $7.3 \times 10^3$ & $0.73$ & $5.3 \times 10^{-3}$ & $2.7 \times 10^{-2}$ & $3.8$\\
$3.0$  & $2048^2$ &  $10^8$ & $0.05$ & $1.7$ & $5.8 \times 10^3$ & $0.6$ & $3.2 \times 10^{-3}$ & $1.2 \times 10^{-2}$ & $4.3$\\
$4.9$  & $512^2$ &  $10^8$ & $0.01$ & $4.2$ & $4.5 \times 10^3$ & $0.45$ & $1.2 \times 10^{-3}$ & $2.9 \times 10^{-3}$ & $1.4$\\
$7.3$ & $8192^2$ &  $10^{10}$ & $0.01$ & $3.4$ & $3.7 \times 10^4$ & $0.37$ & $4.9 \times 10^{-4}$ & $2.8 \times 10^{-3}$ & $4.9$\\
\hline 
\end{tabular}
\label{table:simulation_2D}
\end{center}
\end{table}

\begin{figure}
\begin{center}
\includegraphics[scale=0.8]{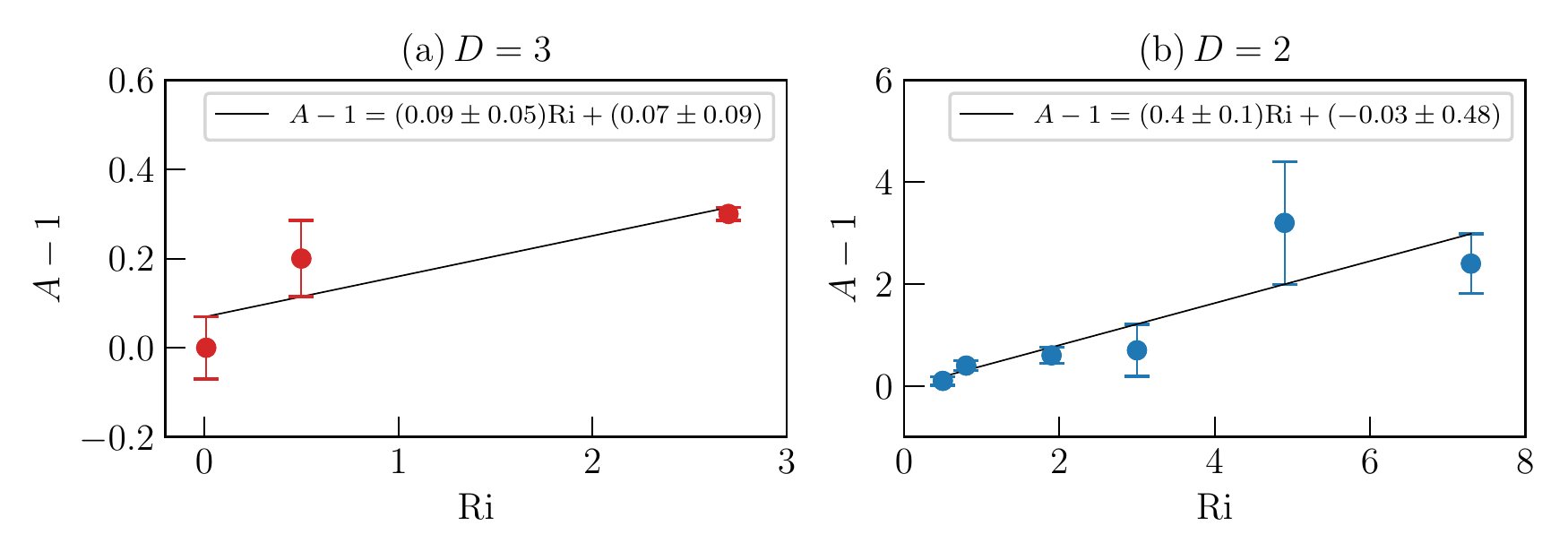}
\end{center}
\setlength{\abovecaptionskip}{0pt}
\caption{Plot of $A-1$ versus the Richardson number $\mathrm{Ri}$ for: (a) three-dimensional ($D=3$) stably stratified turbulence and (b) two-dimensional ($D=2$) stably stratified turbulence. Here $A=  \langle u_1^2 +u_2^2\rangle/(2 \langle u_3^2\rangle) $ in 3D, and $A=  \langle u_1^2 \rangle/ \langle u_3^2\rangle $ in 2D.}
\label{fig:plot}
\end{figure}

In Fig.~\ref{fig:plot}(a, b), we plot $A-1$ versus $\mathrm{Ri}$ for $D=3$ and $D=2$ respectively. We observe  that 
\begin{equation}
A - 1 \propto \mathrm{Ri},
\end{equation}
consistent with  Eq.~(\ref{eq:A_3D}) up to first order.  We  compute the slope of the curves  for $D=2$ and $D=3$  using linear regression and observe that the best fits yields slope $0.09\pm0.05$ for $D=3$, and $0.4\pm0.1$ for $D=2$ with significant error-bar.  Yet, we can argue that our computational estimate is in general agreement with the  theoretical prediction.  The difference could be attributed to various assumptions made in our theoretical formulation.  It is important to remark that we cannot  compare the computational and analytical $A$'s explicitly  due to uncertainties in $\Gamma_1$ and $\sigma_t$. 

Note that for $D=3$, we have only performed three sets of simulations because the three-dimensional computations are very expensive. For $D=2$, we observe a near saturation for $A-1$ for the Richardson number above $10$, as shown in Fig.~\ref{fig:plot}. As discussed by Kumar {\em et al.}~\cite{Kumar:JoT2016}, the two-dimensional stably stratified turbulence above $\mathrm{Ri}>10$ corresponds to the strongly stratified regime and yields vertically sheared horizontal flow (VSHF).

\section{Conclusions}
\label{sec:conc}
We investigated the anisotropy of turbulence in a stably stratified fluid in a regime where the energy spectrum is still Kolmogorov-like. Although the Kolmogorov $-5/3$ law is associated with homogeneous isotropic turbulence, it does not imply that the turbulence in the stratified fluid is isotropic. It simply means that if the horizontal components of the velocity are $v_1$ and $v_2$, and the vertical component is $v_3$, then the energy spectrum $E_\perp(k)$ corresponding to $\int v_\perp^2 d{\bf r} = \int (v_1^2 + v_2^2) d{\bf r}$ scales as $k^{-5/3}$ and the spectrum $E_\parallel(k)$ corresponding to  $\int v_3^2 d{\bf r}$ also scales as $k^{-5/3}$ but the coefficients of $E_\perp(k)$ and $2E_\parallel(k)$ are different. This is what is seen by our calculation of $A= (\int v_\perp^2 d{\bf r})  / (2\int v_\parallel^2 d{\bf r})$. Theoretical computations  using {\em equivalent linearization} technique reveals that $A-1$ increases linearly with the Richardson number. { This is a calculational technique that can be effectively employed in other areas of fluid dynamics and plasma physics, where additionally frictional forces proportional to velocity may be present~\cite{Pometescu:PPCF1998,Lalescu:JP2011,Lalescu:AUC20141}.}

We also employ numerical simulations to verify the aforementioned predictions. It is not possible to compare the computational and analytical $A$'s explicitly  due to uncertainties in $\Gamma_1$ and $\sigma_t$. We observed linear dependence of $A-1$ with $\mathrm{Ri}$ for two-dimensional and three-dimensional direct numerical simulations in the low Richardson number limit.

 As the Richardson number is increased, the anisotropy increases very strongly in the equivalent linearization technique due to the increase in the energy content in the horizontal plane. This is quantitatively correct, but the full nonlinear model shows a saturation at a much lower Richardson number. In this regime, the flow becomes almost two dimensional.  A similar phenomenon happens in two dimensions, where $A-1$ increases linearly with the Richardson number initially for weak to moderate stratification. As Richardson number increases, the full model shows  quite complex behaviour, which, unfortunately, is not captured  in the equivalent linearization approximation.

 \section*{Acknowledgement}
The work of JKB was supported by APS-IUSSTF professorship award and by visiting professorship at IPST, University of Maryland College Park, USA. Several stimulating conversations with Prof Jan Sengers is gratefully acknowledged.  MKV and AK would also like to acknowledge support from the Indian Space Research Organisation, India, for the research grant PLANEX/PHY/2015239 and the Department of Science and Technology, India (INT/RUS/RSF/P-03) and Russian Science Foundation, Russia (RSF-16-41-02012) for the Indo-Russian project. We also gratefully acknowledge the hospitality of the International Centre for Theoretical Sciences (ICTS), Bangalore, India during our visit in ICTS/Prog-buoyant/2017/06, when part of this work was done.

\appendix

\section{Scale invariance of the equations of motion and the form of the effective viscosity}
\label{sec:appA}
{
In this appendix we discuss the conditions under which the equations of motion (Eqs.~(\ref{eq:u_non})-(\ref{eq:theta_non})) will be scale invariant. We start with Eq.~(\ref{eq:u_non}) under the constraint of Eq.~(\ref{eq:inc_non}) and with $\mathrm{Ri}=0$---the usual Navier--Stokes equation. We scale all distances by $l$, time by  $l^s$, and the external random force by  $l^\beta$. Thus we have ${\bf r}=l{\bf r}^\prime$, $\tau=l^s\tau^\prime$, ${\bf u}=l^{1-s}{\bf u}^\prime$,  and  ${\bf f}=l^\beta {\bf f}^\prime$. Using these relations in Eq.~(\ref{eq:u_non}), in terms of the primed variables, Navier--Stokes equation reads (note that the dimensions of the nonlinear term and the pressure gradient term are identical because of the solenoidal constraint of Eq.~(\ref{eq:inc_non}))
\begin{equation}
l^{1-2s}\left[ \frac{\partial \mathbf{u}^\prime}{\partial \tau^\prime} + (\mathbf{u}^\prime \cdot \nabla^\prime) \mathbf{u}^\prime  +  \frac{\nabla^\prime \delta p^\prime}{\rho_0} \right]=  \nu^\prime l^{-1-s} \nabla^{\prime 2} \mathbf{u}^\prime + l^\beta {\bf f}^\prime. 
\label{eq:u_prime}
\end{equation}
To ensure that we have scale invariance, i.e., the above equation in primed variables look exactly like the original Navier--Stokes equation we need
\begin{equation}
\nu^\prime \propto l^{2-s} \qquad  \texttt{and} \qquad  \beta=1-2s. 
\label{eq:nu_prime}
\end{equation}
The above equation implies that the viscosity acquires scale (wave-vector) dependence (Heisenberg's eddy viscosity) and that the correlator $F(k)$ of the random force (see Eq.~(\ref{eq:ff})) will have a special form. From Eq.~(\ref{eq:ff}), we see that the scaling dimension of $F(k)$ is given by $l^{2\beta}l^{D+s}$, which on using Eq.~(\ref{eq:nu_prime}) becomes $l^{D+2-3s}$ and hence
\begin{equation}
F(k) \propto k^{-(D+2-3s)}.
\end{equation}
It remains to find the value of `$s$'. This is done by requiring that the energy transfer rate is scale independent. In the steady state, the rate at which energy is pumped into the system by the stirring force at large length scales equals the rate at which it  is transferred to lower scales and  finally equal to the rate at which it is dissipated at the smallest scales. This rate  $\epsilon$ is equal to the rate at which the space averaged kinetic average changes and is given by
\begin{equation}
\epsilon=\frac{1}{V}\frac{d}{d \tau}\int {u}^{2}({\bf r},\tau)d{\bf r}.
\end{equation}
The scale dependence of the right hand side of the above equation is $l^{2-3s}$ and since Kolmogorov requires this rate to be dimensionless, we see that
\begin{equation}
s=2/3.
\label{eq:s}
\end{equation}
This tells us that the scale dependent viscosity increases with scale as $l^{4/3}$ and the energy spectrum $E(k)$ defined by (in a $D$-dimensional space)
\begin{equation}
E(k) = \langle  {\bf u}(\mathbf{k})\cdot {\bf u}(\mathbf{k}^\prime) \rangle  \delta (\mathbf{k} + \mathbf{k^\prime})/C_Dk^{D-1},
\end{equation}
where $C_D=2\pi^{D/2}/\Gamma(D/2)$ is the surface area of the $D$-dimensional unit sphere, and $\Gamma(x)$ is the gamma function~\cite{Arfken:book}. Using the above results for the scaling dimensions, that of $E(k)$ is easily seen to be
\begin{equation}
E(k) \propto l^{3-2s}.
\end{equation}
Using the value of `$s$' from Eq.~(\ref{eq:s}), we get the Kolmogorov result  $E(k) \propto k^{-5/3}$.  Turning to Eq.~(\ref{eq:s}) now, we see that for $s=2/3$  the random force correlator $F(k)$ scales as  $k^{-D}$. This particular variety of stirring force has been used in all analytical results for the Kolmogorov spectrum starting from a randomly forced Naiver--Stokes equation. What the above analysis demonstrates that it is a requirement of the scale invariance of the equation of motion.

We now include the buoyancy term in the equation of motion which is to say that we work with the full Eq.~(\ref{eq:u_non}).  Consequently we have an additional field  $\theta({\bf r},\tau)$, the temperature fluctuation, and an additional parameter which is the Richardson number $\mathrm{Ri}$.  The scaling dimension of this field is taken to be    $l^\gamma$ and the scaling dimension of $\mathrm{Ri}$ to be $l^\eta$. Performing these scalings in addition to the scalings carried out in  arriving at Eq.~(\ref{eq:u_prime}), we get
\begin{equation}
l^{1-2s}\left[ \frac{\partial \mathbf{u}^\prime}{\partial \tau^\prime} + (\mathbf{u}^\prime \cdot \nabla^\prime) \mathbf{u}^\prime  +  \frac{\nabla^\prime \delta p^\prime}{\rho_0} \right]=  \mathrm{Ri} l^{\gamma+\eta}\theta^\prime+\nu^\prime  l^{-1-s} \nabla^{\prime 2} \mathbf{u}^\prime + l^\beta{\bf f}^\prime. 
\label{eq:theta_prime}
\end{equation}
For scale invariance, we now need in addition to the conditions of Eq.~(\ref{eq:nu_prime}),
\begin{equation}
\gamma+\eta = 1-2s,
\label{eq:gamma}
\end{equation}
which reflects the equal status of inertial and buoyancy forces. The real difference comes in the discussion of the fluxes. For the kinetic energy flux, as before,  the scale factor is  $l^{2-3s}$ while for the thermal flux which is the time derivative of the thermal energy G (see Eq.~(\ref{eq:G})), the scale factor is $l^{2\gamma-s}$. We now have three possibilities:
\begin{enumerate}
\item Both kinetic energy and thermal fluxes are scale invariant.
\item Only the thermal flux is scale invariant.
\item Only the kinetic energy is scale invariant. This occurs at such short length scales as to be unphysical.
\end{enumerate}

We will now discuss the first two cases separately.
\begin{itemize}
\item Case i): Both kinetic energy and thermal fluxes are scale invariant leading to $s=2/3$ and $\gamma=s/2 =1/3$. This leads as before to the energy spectrum $E(k) \propto k^{-5/3}$ and is the Kolmogorov regime. In this case the exponent $\eta$  is seen to be $-2/3$ and that implies that the Richardson number is going to be irrelevant (scales down to zero) as should be for a Kolmogorov like situation.
\item Case ii): In this case only the thermal flux is scale invariant which leads to  $s=2\gamma$. The Richardson number is expected to behave as a dimensionless number and we set $\eta=0$. From Eq.~(\ref{eq:gamma}), we now get   $\gamma=1/5$ leading to $s=2/5$. The kinetic energy flux now falls off as  $k^{-4/5}$. From Eq.~(\ref{eq:gamma}),  the scaling dimension of  the kinetic energy is now $11/5$ and hence one has the Bolgiano--Obukhov scaling of $E(k) \propto k^{-11/5}$. It should be noted that in this Bolgiano--Obukhov regime, the scale invariance of Eq.~(\ref{eq:theta_non}) leads to a scale dependence of the thermal diffusivity in the same fashion as ``eddy" viscosity and the external driving current $h({\bf r},\tau)$  has the same scaling dimension as the external driving in the velocity equation. The relevance of the $z$-component of the velocity in this equation is the indicator of the anisotropy associated with the convective fixed point.
\end{itemize}

To conclude this appendix, we show how the effective viscosity arises from the nonlinear term in Navier--Stokes equation and lays the basis of Eq.~(\ref{eq:u_alpha_k}). Without the external force ${\bf f}$ in Eq.~(\ref{eq:u_non}), which is not necessary for the discussion here, we write the Navier--Stokes equation in momentum space as (defining the Fourier transform: $\psi({\bf k})$ of a function $\psi({\bf r})$ as $\psi({\bf k})=V^{-1/2}\int d{\bf r}e^{i{\bf k}\cdot {\bf r}} \psi({\bf r})$ where $V$ is the volume in which  $\psi({\bf r})$ is defined)
\begin{equation}
\frac{\partial u_i({\bf k})}{\partial t} = -\nu k^2u_i({\bf k})-\sqrt{V} M_{ijl}({\bf k})\int d{\bf p} u_j({\bf p})  u_i({\bf k-p}),
\end{equation}
where
\begin{eqnarray}
M_{ijl} & = & k_iP_{jl}({\bf k})+k_jP_{il}({\bf k})\\
P_{\alpha \beta }({\bf k}) & = & \delta_{\alpha \beta } - \frac{k_\alpha k_\beta}{k^2}
\end{eqnarray}
At the scale $k$, the energy is $E(k)=\langle u_i({\bf k}) u_i({\bf -k}) \rangle$, where the angular bracket indicates the averaging required for a turbulent flow and the total energy $E=\int E(k) 4\pi k^2 dk$. The rate of change of energy as scale $k$ is obtained from Eq.~(\ref{eq:u_non}) as 
\begin{equation}
\frac{\partial E(k)}{\partial t} = -\nu k^2 E(k)-\sqrt{V}M_{ijl} \left \langle \int d{\bf p}u_i({\bf -k})u_j({\bf p})u_l({\bf k-p}) \right \rangle.
\end{equation}
The energy transfer rate across a given wavenumber $\kappa$ is 
\begin{equation}
\epsilon(\kappa)=\int_0^\kappa \frac{\partial E(k)}{\partial t}4\pi k^2 dk
\end{equation}
and is found from Eq.~(\ref{eq:u_12_bar_app}) to be 
\begin{equation}
\epsilon(\kappa)=-\nu \int_0^\kappa \frac{\partial E(k)}{\partial t}4\pi k^4 dk -T(\kappa),
\end{equation}
where
\begin{equation}
T(\kappa) = \sqrt{V}\left \langle \int_0^\kappa d{\bf k}M_{ijl}({\bf k}) u_i({\bf -k}) \int d{\bf p}u_i({\bf -k})u_j({\bf p})u_l({\bf k-p}) \right \rangle
\end{equation}
is the energy transfer across the wavenumber $\kappa$ by the nonlinear term. Making this transfer term look similar to the first term on the right hand side requires us to cast $T(\kappa)$ in the form
\begin{equation}
T(\kappa) - \nu_{eff}(\kappa)\int_0^\kappa E(k)4\pi k^4 dk
\end{equation}
If we look at the RHS of Eq.~(\ref{eq:2u3}) then two factors of velocity, $d{\bf p}$, and a factor of $k^2$ from $d{\bf k}$ constitute dimensionally the part $\int_0^\kappa E(k)4\pi k^4 dk$. The remaining factors, a factor of $k$ from $M_{ijl}$, another from $dk$, $k^{-3/2}$ from $\sqrt{V}$ and $k^{-1/3}k^{-3/2}$ from $\nu(k)$ give the dimensions of $\nu_{eff}(\kappa)$. Since $s=2/3$, we have after converting this coordinate scale to wavenumber space and hence $\nu_{eff}(\kappa)\propto \kappa^{-4/3}$. An identical argument shows that the effective heat diffusion coefficient has a similar $\kappa^{-4/3}$ dependence. }

\section{A calculation of the anisotropy}
\label{sec:appB}
In this appendix, we provide the details behind the result quoted in Eq. (\ref{eq:A_3D}). We begin by dropping the external random ``heat current" $h({\bf r},\tau)$ in Eqs.~(\ref{eq:u_3_bar}) and~(\ref{eq:u_12_bar}) since we want to compare with  simulations that have been carried out with only the random forcing present in the velocity equation. Our starting point then becomes
\begin{equation}
\bar{u_3}^2 = \frac{\left[ \Gamma_2 \left( \Gamma_1 + \Gamma_2 \right) k^{2{s}} + \mathrm{Ri} \sin^2 \xi \right] \langle f_3f_3 \rangle}{2\left( \Gamma_1 + \Gamma_2 \right)k^{s} \left( \Gamma_1 \Gamma_2 k^{2{s}} + \mathrm{Ri} \sin^2 \xi \right)} \label{eq:u_3_bar_app},
\end{equation}
\begin{eqnarray}
\bar{u_1}^2 + \bar{u_2}^2  =  \frac{\langle f_1 f_1 \rangle + \langle f_2 f_2 \rangle}{2 \Gamma_1 k^{s}} + \nonumber \\ 
   \frac{\mathrm{Ri}^2 \cos^2 \xi \sin^2 \xi \left( 2 \Gamma_1 + \Gamma_2 \right) k^{s} \langle f_3 f_3 \rangle}{2 \Gamma_1 \left( \Gamma_1 + \Gamma_2 \right) k^{2{s}} \left( \mathrm{Ri} \sin^2 \xi + \Gamma_1 \Gamma_2 k^{2{s}} \right) \left( 2 \left( \Gamma_1 + \Gamma_2 \right) \Gamma_1 k^{2{s}} + \mathrm{Ri} \sin^2 \xi \right)}. \label{eq:u_12_bar_app}
\end{eqnarray}

Note that  $k^2=k_1^2 + k_2^2 + k_3^2$ in three dimensions ($D=3$). The third direction is along gravity and $k_3$  is discrete---the allowed values in units of `$d$' are $n\pi$. In general we will write $k^2=p^2+n^2\pi^2$  where $\bf p$ is the wave-vector in the horizontal $D$-1 dimensional space.

We see from Eq.~(\ref{eq:ff}) that $\langle f_i f_i \rangle$ is the same for all `i' and will be denoted by $\langle ff \rangle$. For $\mathrm{Ri}=0$ and for all $k$, 
\begin{equation}
\bar{u_3}^2  = \frac{\langle ff \rangle}{2\Gamma_1k^s}=\frac{\bar{u_1}^2 + \bar{u_2}^2 }{2},
\end{equation}
and hence $A=1$ as expected. Expanding  the right hand side of Eq.~(\ref{eq:u_3_bar_app}) in powers of $\mathrm{Ri}$
\begin{eqnarray}
2 \bar{u_3}^2  = \frac{\langle ff \rangle}{\Gamma_1k^s} \frac{\left( 1 + \frac{\mathrm{Ri} \sin^2 \xi}{\Gamma_2  (\Gamma_1 + \Gamma_2)k^{2s}} \right)}{\left( 1 + \frac{\mathrm{Ri} \sin^2 \xi}{ \Gamma_1 \Gamma_2 k^{2s}} \right)} \nonumber \\
 = \frac{\langle ff \rangle}{\Gamma_1k^s} \nonumber \\
\times  \left[ 1 -  \frac{\mathrm{Ri} \sin^2 \xi}{ \Gamma_1 \Gamma_2 k^{2s}} \left( \frac{\Gamma_2}{\Gamma_1 + \Gamma_2} \right) +   \frac{\mathrm{Ri}^2 \sin^4 \xi}{ \Gamma_1^2 \Gamma_2^2 k^{4s}} \left( \frac{\Gamma_2}{\Gamma_1 + \Gamma_2} \right) + \mathcal{O}(\mathrm{Ri}^3) \right]. \label{eq:2u3}
\end{eqnarray}
Similarly for the transverse components we have
\begin{equation}
\bar{u_1}^2 + \bar{u_2}^2 =  \frac{\langle ff \rangle}{\Gamma_1k^s}  \left[ 1 +   \frac{\mathrm{Ri}^2  \cos^2 \xi \sin^2 \xi (2\Gamma_1+\Gamma_2)}{ 4\Gamma_1^2 \Gamma_2 k^{4s} (\Gamma_1 + \Gamma_2)^2} + \mathcal{O}(\mathrm{Ri}^3) \right] \label{eq:u1_u2_appB}.
\end{equation}
To evaluate the global anisotropy factor, the velocity fields have to be integrated over all space. For a $D$-dimensional space what is needed is $\int \bar{u}^2 d^D r$. By Parseval's theorem this integral can be written as 
\begin{equation}
\int \bar{u}^2({\bf r}) d^D r= \int \frac{d^Dk}{(2\pi)^D}\bar{u}^2 ({\bf k}) = \sum_{n=1}^\infty \int \frac{d^{D-1}p}{(2\pi)^{D-1}}\bar{u}(n,{\bf p})^2.
\end{equation}
The scaling requirement that $F(k)\propto k^{-D}$ means $\langle ff \rangle=F_0k^{-D}$, where $F_0$  is a constant, now leads to
\begin{equation}
2\int \bar{u}^2_3 ({\bf r}) d^D r = \frac{F_0}{\Gamma_1}  \left[ I_1 -  \frac{\mathrm{Ri}}{ \Gamma_1 \Gamma_2 } \frac{\Gamma_2}{\Gamma_1 + \Gamma_2}  I_2+   \frac{\mathrm{Ri}^2 }{ \Gamma_1^2 \Gamma_2^2}  \frac{\Gamma_2}{\Gamma_1 + \Gamma_2} I_3 + \mathcal{O}(\mathrm{Ri}^3) \right].
\end{equation}
The integrals $I_1$, $I_2$, and $I_3$ are defined as follows:
\begin{eqnarray}
I_1 & = & \sum_{n=1}^\infty \int \frac{d^{D-1}p}{(2\pi)^{D-1}} \frac{1}{(p^2 + n^2 \pi^2)^{\frac{D+s}{2}}} \nonumber \\
& = &\sum_{n=1}^\infty \frac{1}{(n\pi)^{1+s}}\int \frac{d^{D-1}p}{(2\pi)^{D-1}} \frac{1}{(1+p^2)^{\frac{D+s}{2}}}, \\
I_2 & = & \sum_{n=1}^\infty \int \frac{d^{D-1}p}{(2\pi)^{D-1}} \frac{\sin^2 \xi}{(p^2 + n^2 \pi^2)^{\frac{D+3s}{2}}} \nonumber \\ 
& = & \sum_{n=1}^\infty \frac{1}{(n\pi)^{1+3s}}\int \frac{d^{D-1}p}{(2\pi)^{D-1}} \frac{p^2}{(1+p^2)^{\frac{D+2+3s}{2}}}, \\
I_3 & = & \sum_{n=1}^\infty \frac{1}{(n\pi)^{1+5s}}\int \frac{d^{D-1}p}{(2\pi)^{D-1}} \frac{p^4}{(1+p^2)^{\frac{D+4+5s}{2}}}.
\end{eqnarray}
In evaluating the integrals, we note that each of them carry the factor
\begin{equation}
\frac{\pi^{(D-1)/2}}{2\Gamma\left(\frac{D-1}{2}\right)(2\pi)^{D-1}},
\end{equation}
which we will denote by $C$. We find
\begin{eqnarray}
I_1 & = &C\frac{\zeta(1+s)}{\pi^{1+s}}\frac{\Gamma \left(\frac{1+s}{2}\right) \Gamma \left(\frac{D-1}{2}\right) }{\Gamma \left(\frac{D+s}{2}\right)}, \\
I_2 & = & C\frac{\zeta(1+3s)}{\pi^{1+3s}}\frac{\Gamma \left(\frac{1+3s}{2}\right) \Gamma \left(\frac{D+1}{2}\right) }{\Gamma \left(\frac{D+3s+2}{2}\right)}, \\
I_3 & = & C\frac{\zeta(1+5s)}{\pi^{1+5s}}\frac{\Gamma \left(\frac{1+5s}{2}\right) \Gamma \left(\frac{D+3}{2}\right) }{\Gamma \left(\frac{D+5s+4}{2}\right)}.
\end{eqnarray}
In the above $\zeta(x)$ is the usual Riemann zeta function. As for the horizontal component, as seen from Eq.~(\ref{eq:u1_u2_appB}),
\begin{eqnarray}
&& \int (\bar{u_1}^2 + \bar{u_2}^2) d^D r  \nonumber\\
& = &\frac{F_0}{\Gamma_1} \left[ I_1 +   \frac{\mathrm{Ri}^2   \Gamma_2(2\Gamma_1+\Gamma_2)}{ 4\Gamma_1^2 \Gamma_2^2 (\Gamma_1 + \Gamma_2)^2}I_4 \right].
\end{eqnarray}
The  integral  $I_4$ is seen to be
\begin{eqnarray}
I_4 & = & C\sum_{n=1}^\infty \int_0^\infty \frac{p^{D-1} \sin^2 \xi \cos^2 \xi }{(p^2 + n^2 \pi^2)^{\frac{D+5s}{2}}}dp \nonumber \\
& = & C\frac{\zeta(1+5s)}{\pi^{1+5s}}\frac{ \Gamma \left(\frac{D+1}{2}\right)  \Gamma \left(\frac{3+5s}{2}\right) }{\Gamma \left(\frac{D+4+5s}{2}\right)}. \label{eq:I_4}
\end{eqnarray}
We define $X=\mathrm{Ri}/\Gamma_1\Gamma_2 \pi^{2s}$ and in terms of this parameter, the anisotropy is found to be
\begin{eqnarray}
A & = &\frac{1+X^2\pi^{4s}   \frac{\Gamma_2(2\Gamma_1+\Gamma_2)}{ 4 (\Gamma_1 + \Gamma_2)^2}\frac{I_4}{I_1} + \mathcal{O}(X^3)}{1- X\pi^{2s}\frac{\Gamma_2}{\Gamma_1+\Gamma_2}\frac{I_2}{I_1} + X^2\pi^{4s}\frac{\Gamma_2}{\Gamma_1+\Gamma_2}\frac{I_3}{I_1}+\mathcal{O}(X^3)} \nonumber \\
& = &1+ X\frac{\Gamma_2}{\Gamma_1+\Gamma_2}\frac{I_2\pi^{2s}}{I_1} \nonumber \\
& + & X^2 \left[  \left( \frac{\Gamma_2}{\Gamma_1+\Gamma_2}  \right)^2 \frac{I_2^2\pi^{4s}}{I_1^2} - \frac{\Gamma_2}{\Gamma_1+\Gamma_2} \frac{I_3\pi^{4s}}{I_1} + \frac{\Gamma_2(2\Gamma_1+\Gamma_2)}{(\Gamma_1+\Gamma_2)^2}  \frac{I_4\pi^{4s}}{I_1} \right] \nonumber \\ 
&+& \mathcal{O}(X^3),
\end{eqnarray}    
with
\begin{eqnarray}
\frac{I_2\pi^{2s}}{I_1} & = &\frac{\zeta(1+3s)}{\zeta(1+s)}\frac{\Gamma\left(\frac{D+1}{2}\right) \Gamma\left(\frac{D+s}{2}\right) \Gamma\left(\frac{1+3s}{2}\right)}{\Gamma\left(\frac{D-1}{2}\right) \Gamma\left(\frac{D+2+3s}{2}\right) \Gamma\left(\frac{1+s}{2}\right)}, \\
\frac{I_3\pi^{4s}}{I_1} & = &\frac{\zeta(1+5s)}{\zeta(1+s)}\frac{\Gamma\left(\frac{D+3}{2}\right) \Gamma\left(\frac{D+s}{2}\right) \Gamma\left(\frac{1+5s}{2}\right)}{\Gamma\left(\frac{D-1}{2}\right) \Gamma\left(\frac{D+4+5s}{2}\right) \Gamma\left(\frac{1+s}{2}\right)}, \\
\frac{I_4\pi^{4s}}{I_1} & = &\frac{\zeta(1+5s)}{\zeta(1+s)}\frac{\Gamma\left(\frac{D+1}{2}\right) \Gamma\left(\frac{D+s}{2}\right) \Gamma\left(\frac{3+5s}{2}\right)}{\Gamma\left(\frac{D-1}{2}\right) \Gamma\left(\frac{D+4+5s}{2}\right) \Gamma\left(\frac{1+s}{2}\right)}.
\end{eqnarray}   
In $D=3$, we get 
\begin{eqnarray}
\pi^{2s}\frac{I_2}{I_1}& = &\frac{2\zeta(1+3s)}{3(1+3s)\zeta(1+s)}, \\
\pi^{4s}\frac{I_3}{I_1}& = &\frac{8\zeta(1+5s)}{5(3+5s)(1+5s)\zeta(1+s)}, \\
\pi^{4s}\frac{I_4}{I_1}& =& \frac{2\zeta(1+5s)}{5(3+5s)\zeta(1+s)}.
\end{eqnarray}
We will now calculate the anisotropy using Eq.~(\ref{eq:I_4}) in $D=3$. The isotropic situation in $D=3$ (i.e., the Kolmogorov limit) corresponds to $s=2/3$. In the turbulent state $\Gamma_1 \cong \Gamma_2$  and we get
\begin{equation}
A = 1+X\frac{\pi^{4/3}I_2}{2I_1} + X^2\pi^{8/3}\left[ \frac{3I_4}{4I_1}+\frac{1}{4} \left(\frac{I_2}{I_1}\right)^2-\frac{I_3}{2I_1}\right].\label{eq:aniso_appB}
\end{equation}
Numerical tables provide $\zeta(5/3)\cong2$, $\zeta(3)\cong1.20$, and $\zeta(13/3)\cong1.07$. Thus the coefficient of $X$ in Eq.~(\ref{eq:aniso_appB}) is $0.067$ and the coefficient of $X^2$ works out to be $0.016$. We find the anisotropy in three dimensions to be given by
\begin{equation}
A=1+0.067X+0.016X^2+\mathcal{O}(X^3), \label{eq:aniso_appB_2}
\end{equation}
where $X=\mathrm{Ri}/\Gamma_1\Gamma_2\pi^{4/3}$. The relaxation rates will be largest at the smallest wavenumber and for the geometry considered the smallest wavenumber has the magnitude $\pi$ in units of $1/d$.  The denominator of $X$ is the product of the largest viscous and thermal relaxation rates and can be of $\mathcal{O}(1)$ but certainly less than unity since the bare values are much smaller than unity.


\begin{thebibliography}{10}

\bibitem{Arfken:book}
G.~Arfken and H.~Weber.
\newblock {\em Mathematical methods for physicists}.
\newblock Elsevier Acad. Press, 2008.

\bibitem{Bartello:JFM2013}
P.~Bartello and S.~M. Tobias.
\newblock {Sensitivity of stratified turbulence to the buoyancy Reynolds
  number}.
\newblock {\em J. Fluid Mech.}, 725:1--22, June 2013.

\bibitem{Bhattacharjee:PLA2015}
J.~K. Bhattacharjee.
\newblock {Kolmogorov argument for the scaling of the energy spectrum in a
  stratified fluid}.
\newblock {\em Phys. Lett. A}, 379(7):696--699, Mar. 2015.

\bibitem{Bolgiano:JGR1959}
R.~Bolgiano.
\newblock {Turbulent spectra in a stably stratified atmosphere}.
\newblock {\em J. Geophys. Res.}, 64(12):2226--2229, 1959.

\bibitem{Brethouwer:JFM2007}
G.~Brethouwer, P.~Billant, P.~Billant, E.~Lindborg, and J.-M. Chomaz.
\newblock {Scaling analysis and simulation of strongly stratified turbulent
  flows}.
\newblock {\em J. Fluid Mech.}, 585:343--368, Aug. 2007.

\bibitem{Chandrasekhar:PRSA1949}
S.~Chandrasekhar.
\newblock On {H}eisenberg{\textquoteright}s elementary theory of turbulence.
\newblock {\em Proc. R. Soc. A}, 200(1060):20--33, 1949.

\bibitem{Chandrasekhar:book:Instability}
S.~Chandrasekhar.
\newblock {\em {Hydrodynamic and Hydromagnetic Stability}}.
\newblock Clarendon Press, 1968.

\bibitem{Chatterjee:JPDC17}
A.~G. Chatterjee, M.~K. Verma, A.~Kumar, R.~Samtaney, B.~Hadri, and R.~Khurram.
\newblock {Scaling of a Fast Fourier Transform and a pseudo-spectral fluid
  solver up to 196608 cores}.
\newblock {\em J. Parallel Distrib. Comput.}, 113:77--91, 2017.

\bibitem{Davidson:book:TurbulenceRotating}
P.~A. Davidson.
\newblock {\em {Turbulence in Rotating, Stratified and Electrically Conducting
  Fluids}}.
\newblock Cambdrige University Press, Cambdrige, 2013.

\bibitem{deBruynKops:JFM2015}
S.~M. de~Bruyn~Kops.
\newblock {Classical scaling and intermittency in strongly stratified
  Boussinesq turbulence}.
\newblock {\em J. Fluid Mech.}, 775:436--463, 2015.

\bibitem{Heisenberg:PRA1948}
W.~Heisenberg.
\newblock On the theory of statistical and isotropic turbulence.
\newblock {\em Proc. R. Soc. A}, 195(1042):402--406, 1948.

\bibitem{Ishida:PF2007}
T.~Ishida and Y.~Kaneda.
\newblock {Small-scale anisotropy in magnetohydrodynamic turbulence under a
  strong uniform magnetic field}.
\newblock {\em Phys. Fluids}, 19(7):075104, 2007.

\bibitem{Kaneda:NJP2004}
Y.~Kaneda and K.~Yoshida.
\newblock {Small-scale anisotropy in stably stratified turbulence}.
\newblock {\em New J. Phys.}, 6:34, Mar. 2004.

\bibitem{Kimura:JFM1996}
Y.~Kimura and J.~R. Herring.
\newblock {Diffusion in stably stratified turbulence}.
\newblock {\em J. Fluid Mech.}, 328:253--269, 1996.

\bibitem{Kimura:JFM2012}
Y.~Kimura and J.~R. Herring.
\newblock {Energy spectra of stably stratified turbulence}.
\newblock {\em J. Fluid Mech.}, 698:19--50, 2012.

\bibitem{Kolmogorov:DANS1941Structure}
A.~N. Kolmogorov.
\newblock {The local structure of turbulence in incompressible viscous fluid
  for very large Reynolds numbers}.
\newblock {\em Dokl Acad Nauk SSSR}, 30(4):301--305, 1941.

\bibitem{Kumar:PRE2014}
A.~Kumar, A.~G. Chatterjee, and M.~K. Verma.
\newblock {Energy spectrum of buoyancy-driven turbulence}.
\newblock {\em Phys. Rev. E}, 90(2):023016, Aug. 2014.

\bibitem{Kumar:JoT2016}
A.~Kumar, M.~K. Verma, and J.~Sukhatme.
\newblock {Phenomenology of two-dimensional stably stratified turbulence under
  large-scale forcing}.
\newblock {\em J. Turbul.}, 18(3):219--239, 2017.

\bibitem{Lalescu:JP2011}
C.~C. Lalescu, D.~Carati, M.~Negrea, and I.~Petrisor.
\newblock Particle transport in incompressible {MHD} kolmogorov flow.
\newblock {\em Journal of Physics: Conference Series}, 333:012010, 2011.

\bibitem{Lalescu:AUC20141}
C.~C. Lalescu, I.~Petrisor, M.~Negrea, and I.~Petrisor.
\newblock Test particles transport in two-dimensional turbulent plasma.
\newblock {\em Annals of the University of Craiova, Physics}, 24:97--103, 2014.

\bibitem{Lindborg:JFM2006}
E.~Lindborg.
\newblock {The energy cascade in a strongly stratified fluid}.
\newblock {\em J. Fluid Mech.}, 550:207--242, Mar. 2006.

\bibitem{Obukhov:DANS1959}
A.~M. Obukhov.
\newblock {On influence of buoyancy forces on the structure of temperature
  field in a turbulent flow}.
\newblock {\em Dokl Acad Nauk SSSR}, 125:1246, 1959.

\bibitem{Pometescu:PPCF1998}
N.~Pometescu, M.~Negrea, and P.~Rotaru.
\newblock The anomalous particle flux induced by electromagnetic turbulence.
\newblock {\em Plasma Phys. Control. Fusion}, 40(7):1383--1398, 1998.

\bibitem{PRAUD:JFM2005}
O.~Praud, A.~M. Fincham, and J.~Sommeria.
\newblock {Decaying grid turbulence in a strongly stratified fluid}.
\newblock {\em J. Fluid Mech.}, 522:1--33, Jan. 2005.

\bibitem{Rorai:PRE2015}
C.~Rorai, P.~D. Mininni, and A.~G. Pouquet.
\newblock {Stably stratified turbulence in the presence of large-scale
  forcing}.
\newblock {\em Phys. Rev. E}, 92(1):013003, 2015.

\bibitem{Seychelles:PRL2008}
F.~Seychelles, Y.~Amarouchene, M.~Bessafi, and H.~Kellay.
\newblock {Thermal convection and emergence of Isolated vortices in soap
  bubbles}.
\newblock {\em Phys. Rev. Lett.}, 100(14):144501--4, Apr. 2008.

\bibitem{Vallgren:PRL2011}
A.~Vallgren, E.~Deusebio, and E.~Lindborg.
\newblock {Possible explanation of the atmospheric kinetic and potential energy
  spectra}.
\newblock {\em Phys. Rev. Lett.}, 107(26):268501, Dec. 2011.

\bibitem{Verma:book}
M.~Verma.
\newblock {\em Physics of Buoyant Flows: From Instabilities to Turbulence}.
\newblock World Scientific, Singapore, 2018.

\bibitem{Verma:ROPP2017}
M.~K. Verma.
\newblock {Anisotropy in Quasi-Static Magnetohydrodynamic Turbulence}.
\newblock {\em Rep. Prog. Phys.}, 80(8):087001--32, May 2017.

\bibitem{verma:NJP2017}
M.~K. Verma, A.~Kumar, and A.~Pandey.
\newblock {Phenomenology of buoyancy-driven turbulence: recent results}.
\newblock {\em New J. Phys.}, 19:025012, 2017.

\bibitem{WAITE:JFM19992}
M.~L. Waite and P.~Bartello.
\newblock {Stratified turbulence dominated by vortical motion}.
\newblock {\em J. Fluid Mech.}, 517:281--308, 1999.

\bibitem{Zhang:PRL2005}
J.~Zhang, X.~L. Wu, and K.-Q. Xia.
\newblock {Density fluctuations in strongly stratified two-dimensional
  turbulence}.
\newblock {\em Phys. Rev. Lett.}, 94(17):174503, May 2005.

\end{thebibliography}

\end{document}